\pdfoutput=1
\RequirePackage{ifpdf}
\ifpdf %We are running pdfTeX in pdf mode
\documentclass[pdftex]{sigma}
\else
\documentclass{sigma}
\fi

\numberwithin{equation}{section}

\newtheorem{Theorem}{Theorem}[section]
\newtheorem{Lemma}[Theorem]{Lemma}
\newtheorem{Proposition}[Theorem]{Proposition}
{ \theoremstyle{definition}
\newtheorem{Example}[Theorem]{Example}
\newtheorem{Remark}[Theorem]{Remark} }

\begin{document}

\newcommand{\arXivNumber}{1408.3019}

\allowdisplaybreaks

\renewcommand{\PaperNumber}{009}

\FirstPageHeading

\ShortArticleName{Lagrangian Reduction on Homogeneous Spaces with Advected Parameters}

\ArticleName{Lagrangian Reduction on Homogeneous Spaces\\
with Advected Parameters}

\Author{Cornelia VIZMAN}

\AuthorNameForHeading{C.~Vizman}

\Address{Department of Mathematics, West University of Timi\c{s}oara, Romania}
\Email{\href{mailto:vizman@math.uvt.ro}{vizman@math.uvt.ro}}

\ArticleDates{Received August 14, 2014, in f\/inal form January 22, 2015; Published online January 29, 2015}

\Abstract{We study the Euler--Lagrange equations for a~parameter dependent $G$-invariant Lagrangian on
a~homogeneous $G$-space.
We consider the pullback of the parameter dependent Lagrangian to the Lie group~$G$, emphasizing the special invariance
properties of the associated Euler--Poincar\'e equations with advected parameters.}


\Classification{53D17; 53D20; 37K65; 58D05; 58D10}

\section{Introduction}
%\label{s1}

The Euler--Poincar\'e (EP) equations arise via reduction of the variational principle for a~right $G$-invariant
Lagrangian $L:TG\to\mathbb R$.
With a~restricted class of variations, the extremals of the integral of the reduced Lagrangian $\ell:\mathfrak
g\to\mathbb R$ correspond to extremals of the original variational problem for~$L$~\cite{MR}.
The EP equations are written for the right logarithmic derivative $\xi=\dot g g^{-1}=\delta^rg$ of the curve~$g$ in~$G$
as
\begin{gather}
\label{epintro}
\frac{d}{dt}\frac{\delta\ell}{\delta \xi}+\operatorname{ad}^*_\xi\frac{\delta\ell}{\delta\xi}=0.
\end{gather}
Here $\delta \ell/\delta\xi$ denotes the functional derivative of~$\ell$, which depends on the choice of a~space
$\mathfrak{g}^*$ in duality with $\mathfrak{g}$, that is $(\frac{\delta
\ell}{\delta\xi},\eta)=\left.\frac{d}{dt}\right|_{t=0}\ell(\xi+t\eta)$ for all $\eta\in\mathfrak{g}$.

The case of a~parameter dependent $G$-invariant Lagrangian $L:TG\times V^*\to\mathbb R$ is studied in~\cite{CHMR}.
The parameter space $V^*$ is a~linear representation space of the Lie group~$G$ and the associated EP equations include
an advection equation for the parameter.
These EP equations with advected parameters are applied to continuum theories in~\cite{HMR}.
To integrate complex f\/luids in this setting, the case of an af\/f\/ine $G$-action on the parameter space $V^*$ is treated
in~\cite{GBR}.
The more general case when the parameter space is a~smooth manifold~$M$ acted on by~$G$ is considered in~\cite{GBT},
applied there to nematic particles.
The reduced equations, called EP equations for symmetry breaking, written for the reduced Lagrangian $\ell:\mathfrak
g\times M\to\mathbb R$, involve the cotangent momentum map $J:T^*M\to\mathfrak g^*$:
\begin{gather*}
\frac{d}{dt}\frac{\delta\ell}{\delta\xi}+\operatorname{ad}^*_\xi\frac{\delta
\ell}{\delta\xi}=J\left(\frac{\delta\ell}{\delta m}\right).
\end{gather*}

In this paper we generalize the Lagrangian reduction with avdected parameters from the Lie group setting to the
homogeneous space setting.
We use the approach from~\cite{TV} that features the special invariance properties of the reduced equations written for
the pullback Lagrangian to the Lie group~$G$.
Starting with a~right $G$-invariant Lagrangian $\bar L:T(G/H)\to\mathbb R$, the reduced Lagrangian $\ell:\mathfrak
g\to\mathbb R$ coming from its pullback to $TG$ must be invariant under the adjoint action of the Lie subgroup~$H$ and
under the addition action of its Lie algebra $\mathfrak h$.
As a~consequence, the EP equations~\eqref{epintro} are now invariant under the following action of the group
$C^{\infty}(I,H)$ on~$C^{\infty}(I,\mathfrak g)$:
\begin{gather*}
h\cdot \xi=\operatorname{Ad}_h\xi+\delta^rh.
\end{gather*}

The geodesic equations for invariant Riemannian metrics on Lie groups (Euler equations) correspond to reduced
Lagrangians~$\ell$ that are quadratic; a~famous example is the ideal f\/luid f\/low as geodesic equations on the group of
volume preserving dif\/feomorphisms~\cite{Arnold} (more geodesic equations on dif\/feomorphism groups can be found for
instance in~\cite{Vizman}).
The extension of Euler equations from Lie groups to homogeneous spaces is done in~\cite{KM}.

The plan of the paper is the following.
In Section~\ref{s2} we review a~kind of logarithmic derivative for homogeneous spaces.
Then we consider parameter dependent $G$-invariant Lagrangians on~$G/H$.
We treat the case of a~linear action on the parameter space in Section~\ref{s3}, where a~multidimensional Hunter--Saxton
equation with advected parameter is obtained.
We devote Section~\ref{s5} to the EP equations for symmetry breaking, obtained for general actions on arbitrary
parameter spaces.
We obtain the af\/f\/ine EP equations as a~special case.
The examples are mainly on inf\/inite dimensional homogeneous spaces, such as $\operatorname{Dif\/f}(S^1)/S^1$,
$\operatorname{Dif\/f}(M)/\operatorname{Dif\/f}_{\operatorname{vol}}(M)$,
$\operatorname{Dif\/f}(M)/\operatorname{Dif\/f}_{\operatorname{iso}}(M)$, and $C^{\infty}(M,G)/G$.

\section{Logarithmic derivative and EP equations}
\label{s2}

The Euler--Lagrange (EL) equations associated to right invariant Lagrangians on Lie groups lead to the Euler--Poincar\'e
(EP) equations involving reduced Lagrangians and written for right logarithmic derivatives of curves in the Lie group.
In this section we recall how this works in homogeneous spaces of right cosets, following~\cite{TV}.

Given a~smooth curve $\bar g:I\to G/H$, we compare the right logarithmic derivatives of two smooth lifts $g_1,g_2:I\to
G$ of $\bar g$.
Because there exists a~smooth curve $h:I\to H$ such that $g_2=hg_1$, the logarithmic derivative $\delta^r
g_2=\operatorname{Ad}_h\delta^rg_1+\delta^rh$ is obtained from $\delta^r g_1$ via the left action of the group
$C^{\infty}(I,H)$ on $C^{\infty}(I,\mathfrak g)$:
\begin{gather}
\label{leftaction}
h\cdot \xi=\operatorname{Ad}_h\xi+\delta^rh.
\end{gather}
Hence the right logarithmic derivative for homogeneous spaces is multivalued:
\begin{gather}
\label{logderhom}
\bar\delta^r: \ C^{\infty}(I,G/H)\to C^{\infty}(I,\mathfrak g)/C^{\infty}(I,H),
\qquad
\bar\delta^r\bar g=C^{\infty}(I,H)\cdot\delta^r g,
\end{gather}
where~$g$ is any lift of $\bar g$.

The tangent bundle $TG$ of a~Lie group~$G$ carries a~natural group multiplication.
In the right trivialization $TG\cong \mathfrak g\times G$ the multiplication becomes
$(\eta,h)(\xi,g)=(\operatorname{Ad}_h\xi+\eta,hg)$.
Given a~Lie subgroup~$H$ of~$G$, its tangent bundle $TH$ is a~subgroup of $TG$.
Let $\pi:G\to G/H$ denote the canonical projection.
Then the   surjective submersion $T\pi:TG\to T(G/H)$ is constant on right $TH$-cosets of $TG$ and descends to
a~canonical dif\/feomorphism between $TG/TH$ and $T(G/H)$.

The following are equivalent data: right $G$-invariant Lagrangian $\bar L$ on $T(G/H)$, left $TH$-invariant and
right $G$-invariant Lagrangian~$L$ on $TG$, as well as reduced Lagrangian~$\ell$ on $\mathfrak g$ that is both
$\mathfrak h$-invariant and $\operatorname{Ad}(H)$-invariant.
The relation between the Lagrangians is $L=\bar L\circ T\pi$ and we call~$L$ the pullback of $\bar L$.

\begin{Proposition}[\cite{TV}]
\label{elhomo}
If the reduced Lagrangian $\ell:\mathfrak g\to\mathbb R$ is $\mathfrak h$--invariant and
$\operatorname{Ad}(H)$-invariant, i.e.\ $\ell(\operatorname{Ad}_h\xi+\eta)=\ell(\xi)$ for all $h\in H$ and
$\eta\in\mathfrak h$, then the EP equations
\begin{gather}
\label{ep}
\frac{d}{dt}\frac{\delta\ell}{\delta \xi}+\operatorname{ad}^*_\xi\frac{\delta\ell}{\delta\xi}=0
\end{gather}
are $C^{\infty}(I,H)$-invariant for the action~\eqref{leftaction}.
\end{Proposition}

In other words, equation~\eqref{ep} can be seen as an equation for $C^{\infty}(I,H)$-orbits in $C^{\infty}(I,\mathfrak g)$,
i.e.\ an equation for right logarithmic derivatives~\eqref{logderhom} of curves in the homogeneous space, so
Proposition~\ref{elhomo} can be reformulated as:

\begin{Proposition}[\cite{TV}]
A~solution of the EL equation for a~right $G$-invariant Lagrangian $\bar L:T(G/H)\to\mathbb R$ is a~curve in
$G/H$ such that the logarithmic derivative of one of its lifts to~$G$ satisfies the EP equation~\eqref{ep}, with~$\ell$
the reduced Lagrangian of the pullback $L:TG\to\mathbb R$.
\end{Proposition}

This proposition admits a~generalization to Lagrangians that are not necessarily right $G$-invariant.
First we note the following property of a~left $TH$-invariant Lagrangian $L:TG\to\mathbb R$: if the curve~$g$ in~$G$ is
a~solution of the corresponding EL equation, then the curve $hg$ is also a~solution of the EL equation, for any smooth
curve~$h$ in~$H$.
Indeed, each variation $g_\varepsilon$ of~$g$ with f\/ixed endpoints corresponds to a~variation $(\varepsilon,t)\mapsto
h(t)g_\varepsilon(t)$ of $hg$ with f\/ixed endpoints.

\begin{Proposition}
Let $L:TG\to\mathbb R$ be the pullback of the Lagrangian $\bar L:T(G/H)\to\mathbb R$ $($i.e.~$L$ is left $TH$-invariant$)$.
Then the following assertions hold:
\begin{enumerate}\itemsep=0pt
\item[$(i)$] If the curve~$g$ in~$G$ is a~solution of the EL equation for~$L$, then it descends to the solution $\bar g=\pi\circ
g$ of the EL equation for $\bar L$.
\item[$(ii)$] If the curve $\bar g$ in $G/H$ is a~solution of the EL equation for $\bar L$, then any lift~$g$ of $\bar g$ is
a~solution of the EL equation for~$L$.
\end{enumerate}
\end{Proposition}

\begin{proof}
Let~$g$ be a~solution of the EL equation for~$L$ and $\bar g=\pi\circ g$.
An arbitrary varia\-tion~$\bar g_\varepsilon$ of $\bar g$ in $G/H$ with f\/ixed endpoints can be lifted to a~variation
$g_\varepsilon$ of~$g$ in~$G$, but it doesn't necessarily have f\/ixed endpoints.
It only satisf\/ies $g_\varepsilon(0)\in Hg(0)$ and $g_\varepsilon(1)\in Hg(1)$.
We can achieve $g_\varepsilon(0)=g(0)$ by multiplying $g_\varepsilon(t)$ with $g(0)g_\varepsilon(0)^{-1}$ from the left.
Moreover, we achieve $g_\varepsilon(1)=g(1)$ by multiplying the new variation $g_\varepsilon(t)$ with
$g(1)g_{\varepsilon t}(1)^{-1}$ from the left.
Now, using also the identity $L=\bar L\circ T\pi$, we get
\begin{gather*}
\frac{d}{d\varepsilon}\Big|_0\int\bar L(\bar g_\varepsilon(t),\dot{\bar {g_\varepsilon}}(t))dt
=\frac{d}{d\varepsilon}\Big|_0\int L(g_\varepsilon(t),\dot{g_\varepsilon}(t))dt =0,
\end{gather*}
so that $\bar g$ is a~solution of the EL equation for $\bar L$.
This proves the f\/irst assertion.

The second assertion is straightforward, since a~variation of~$g$ in~$G$ with f\/ixed endpoints always descends to
a~variation in $G/H$ with f\/ixed endpoints.
\end{proof}

A special case is the geodesic equation for a~right $G$--invariant Riemannian metric on $G/H$, i.e.\ Euler equation on
homogeneous spaces~\cite{KM}.
The next examples are both of this type.

\begin{Example}[\cite{KLMP}]
\label{ex:HS}
Let $(M,\mu)$ be a~volume manifold.
The homogeneous space of right cosets $\operatorname{Dif\/f}(M)/\operatorname{Dif\/f}_{\operatorname{vol}}(M)$ is the space
of normalized volume forms.
The right invariant metric on $\operatorname{Dif\/f}(M)/\operatorname{Dif\/f}_{\operatorname{vol}}(M)$ induced by the
degenerate $\dot H^1$ inner product on $\mathfrak X(M)$
\begin{gather*}
\langle u,v\rangle=\int_M\operatorname{div} u\operatorname{div} v\, \mu
\end{gather*}
is isometric to the standard $L^2$ metric on an open subset of the sphere of radius $2\sqrt{\operatorname{vol}(M)}$ in
the Hilbert space $L^2(M)$.
The isometry is
\begin{gather*}
\bar\varphi\in\operatorname{Dif\/f}(M)/\operatorname{Dif\/f}_{\operatorname{vol}}(M)\mapsto
2\sqrt{\operatorname{Jac}(\varphi)}\in L^2(M),
\end{gather*}
where the Jacobian of $\varphi\in\operatorname{Dif\/f}(M)$ is computed w.r.t.~$\mu$, i.e.\
$\varphi^*\mu=\operatorname{Jac}(\varphi)\mu$.

The geodesic equation is the {multidimensional Hunter--Saxton} equation
\begin{gather}
\label{hsm}
\partial_t d(\operatorname{div} u)+d L_u (\operatorname{div} u)+(\operatorname{div} u) d(\operatorname{div} u)=0.
\end{gather}
The reduced Lagrangian $ \ell(u)=\frac12\int_M(\operatorname{div} u)^2\mu$ on $\mathfrak X(M)$ has the required
$\operatorname{Dif\/f}_{\operatorname{vol}}(M)$- and $\mathfrak X_{\operatorname{vol}}(M)$-invariance properties.
For $M=S^1$ one gets the Hunter--Saxton equation as geodesic equation on $\operatorname{Dif\/f}(S^1)/S^1$.

The left action~\eqref{leftaction} involved in the def\/inition of the right logarithmic derivative $\bar\delta^r$ on the
ho\-mo\-ge\-neous space $\operatorname{Dif\/f}(M)/\operatorname{Dif\/f}_{\operatorname{vol}}(M)$ is the action of the group
$C^{\infty}(I{,}\operatorname{Dif\/f}_{\operatorname{vol}}(M))$ on $C^{\infty}(I{,}\mathfrak X(M))$ given~by
\begin{gather}
\label{acte}
(\psi\cdot u)(t)=\big(\psi(t)^{-1}\big)^*u(t)+\delta^r\psi(t)\in\mathfrak X(M)
\end{gather}
since $\operatorname{Ad}_\psi u=(\psi^{-1})^*u$.
Here the right logarithmic derivative $\delta^r\psi=\partial_t\psi\circ\psi^{-1}$ is the time dependent (divergence
free) vector f\/ield induced by the (volume preserving) isotopy $\psi(t)$.
By Proposition~\ref{elhomo} the Hunter--Saxton equation~\eqref{hsm} is invariant under the action~\eqref{acte}.
This can be checked also by a~direct computation.
\end{Example}

\begin{Example}[\cite{KLMP}]
\label{RM}
Let $(M,g)$ be a~Riemannian manifold and $\operatorname{Dif\/f}_{\operatorname{iso}}(M)$ its group of isometries.
The homogeneous space of right cosets $\operatorname{Dif\/f}(M)/\operatorname{Dif\/f}_{\operatorname{iso}}(M)$ admits
a~right invariant metric induced by the degenerate inner product on $\mathfrak X(M)$
\begin{gather*}
\langle u,v\rangle=\int_M(L_ug,L_vg)\,\mu
=\int_M \big(2\big(du^\flat,dv^\flat\big)+4\big(\delta u^\flat,\delta v^\flat\big)-4\operatorname{Ric}(u,v)\big)\mu.
\end{gather*}
The reduced Lagrangian $\ell(u)=\frac12\int_M|L_ug|^2\mu$ on $\mathfrak X(M)$ has the required
$\operatorname{Dif\/f}_{\operatorname{vol}}(M)$- and \linebreak $\mathfrak X_{\operatorname{vol}}(M)$-invariance properties.
The associated EP equation
\begin{gather*}
4d\delta u^\flat_t+2\delta du^\flat_t-4\operatorname{Ric}(u_t)  +(\operatorname{div} u)
\big(4d\delta u^\flat+2\delta du^\flat-4\operatorname{Ric}(u)\big)
\\
\hphantom{4d\delta u^\flat_t}
{}+L_u\big(4d\delta u^\flat+2\delta du^\flat-4\operatorname{Ric}(u)\big)=0
\end{gather*}
is invariant under the action~\eqref{acte} of $C^{\infty}(I,\operatorname{Dif\/f}_{\operatorname{iso}}(M))$ on
$C^{\infty}(I,\mathfrak X(M))$.
\end{Example}

\section{EP equations with advected parameters}
\label{s3}

Now we look at parameter dependent Lagrangians.
First we treat the Lie group case, follo\-wing~\cite{CHMR}, then we pass to homogeneous spaces.

\subsection{The case of Lie groups}

We consider a~linear right action~$\rho$ of the Lie group~$G$ on the vector space~$V$ and its dual left action $\rho^*$
on $V^*$.
The corresponding Lie algebra actions on~$V$ and $V^*$ are $\tfrac{d}{dt}\big|_0\rho_{\exp(t\xi)}(v)=v\xi$ and
$\tfrac{d}{dt}\big|_0\rho^*_{\exp(t\xi)}(a)=\xi a$.
If $\xi(t)=\delta^rg(t)$, then $a(t)=\rho^*_{g(t)}(a_0)$ is the unique solution of the dif\/ferential equation with
time-dependent coef\/f\/icients $\dot{a}=\xi a$, $a(0)=a_0$.
The diamond operation $\diamond:V\times V^*\to\mathfrak g^*$ is given~by
\begin{gather}
\label{dia}
\langle v\diamond a,\xi\rangle:=\langle \xi a,v\rangle,
\qquad
\text{for all}
\quad
\xi\in\mathfrak g.
\end{gather}

A right $G$-invariant Lagrangian $L:TG\times V^*\rightarrow\mathbb{R}$ (including the linear action on the parameter
space in the second argument) has a~reduced Lagrangian {$\ell:\mathfrak{g}\times V^*\rightarrow\mathbb{R}$} so that
\begin{gather*}
\ell\big(v_gg^{-1},\rho^*_{g}(a)\big)=L(v_g,a),
\qquad
v_g\in T_gG.
\end{gather*}
For f\/ixed $a_0\in V^*$ the Lagrangian $L_{a_0}:TG\rightarrow\mathbb{R}$ is right invariant only under the isotropy
subgroup $G_{a_0}$ of $a_0\in V^*$.

\begin{Theorem}[\cite{CHMR}]
\label{lini}
The EL equations for $L_{a_0}$ on~$G$ given by Hamilton's variational principle
\begin{gather*}
\delta\int_{t_1}^{t_2}L_{a_0}(g(t),\dot{g}(t))dt=0
\end{gather*}
can be expressed as EP equations on $\mathfrak g\times V^*$ with advected parameter:
\begin{gather}
\label{rightep}
{\frac{d}{dt}\frac{\delta\ell}{\delta\xi}+\operatorname{ad}^*_\xi\frac{\delta \ell}{\delta\xi}=\frac{\delta\ell}{\delta
a}\diamond a},
\qquad
\dot{a}=\xi a
\end{gather}
for the reduced Lagrangian~$\ell$.
\end{Theorem}

The main examples are the heavy top and the ideal compressible f\/luid.
For the heavy top $G=\operatorname{SO}(3)$ and the parameter $\Gamma\in V^*=\mathbb R^3$ is the unit vector in the
gravity direction in body representation.
For the ideal compressible f\/luid $G=\operatorname{Dif\/f}(M)$, with~$M$ a~Riemannian manifold, and the parameter $\rho\in
V^*= C^{\infty}(M)^*$ is the f\/luid density in spatial representation.
The reduced Lagrangians are $\ell(\Omega,\Gamma)=\frac12 \operatorname{I}\Omega\cdot\Omega-\Gamma\cdot\lambda$ for
$\Omega\in\mathfrak{so}(3)=\mathbb R^3$ in the f\/irst example, and $\ell(u,\rho)=\frac12\int_M|u|^2\rho$ for
$u\in\mathfrak X(M)$ in the second one.

\subsection{The case of homogeneous spaces}

Let $L:TG\times V^*\to\mathbb R$ be the pull-back of a $G$-invariant Lagrangian $\bar L:T(G/H)\times V^*\to\mathbb R$,
hence~$L$ is left $TH$-invariant and right $G$-invariant.
If $\ell:\mathfrak g\times V^*\to\mathbb R$ is the reduced Lagrangian, then
\begin{gather*}
\ell(\xi,a)=L(\xi,a)=L(v_h\xi,a) =\ell\big({\operatorname{Ad}_h\xi}+v_hh^{-1},\rho^*_{h}(a)\big),
\qquad
v_h\in T_hH.
\end{gather*}
This proves the next proposition.

\begin{Proposition}
%\label{lagrinv}
The reduced Lagrangian $\ell:\mathfrak g\times V^*\to\mathbb R$ associated to the pullback of a~parameter dependent
right $G$-invariant Lagrangian on $G/H$ is~$H$- and $\mathfrak h$-invariant:
\begin{gather}
\label{inv}
\ell(\operatorname{Ad}_h\xi+\eta,\rho^*_h(a))=\ell(\xi,a),
\qquad
h\in H,
\qquad
\eta\in\mathfrak h.
\end{gather}
\end{Proposition}

\begin{Lemma}
\label{barbar}
The functional derivatives of the reduced Lagrangian $\ell:\mathfrak g\times V^*\to\mathbb R$ that has the invariance
property~\eqref{inv} are equivariant:
\begin{gather*}
\frac{\delta\ell}{\delta\xi}(\operatorname{Ad}_h\xi+\eta,\rho^*_h(a))=\operatorname{Ad}^*_{h^{-1}}\frac{\delta
\ell}{\delta \xi}(\xi,a)
\qquad
\text{and}
\qquad
\frac{\delta\ell}{\delta a}(\operatorname{Ad}_h\xi+\eta,\rho^*_h(a))=\rho_{h^{-1}}\frac{\delta \ell}{\delta a}(\xi,a).
\end{gather*}
\end{Lemma}

\begin{proof}
We compute for $\zeta\in\mathfrak g$:
\begin{gather*}
\left(\frac{\delta\ell}{\delta\xi}(\operatorname{Ad}_h\xi+\eta,\rho^*_h(a)),\zeta\right)
 =\frac{d}{dt}\Big|_0\ell(\operatorname{Ad}_h\xi+\eta+t\zeta,\rho^*_h(a))
=\frac{d}{dt}\Big|_0\ell(\xi+t\operatorname{Ad}^*_{h^{-1}},a)
\\
\hphantom{\left(\frac{\delta\ell}{\delta\xi}(\operatorname{Ad}_h\xi+\eta,\rho^*_h(a)),\zeta\right)}{}
 =\left(\operatorname{Ad}^*_{h^{-1}}\frac{\delta\ell}{\delta\xi}(\xi,a),\zeta\right).
\end{gather*}
Similarly we get that
\begin{gather*}
\left(\frac{\delta\ell}{\delta a}(\operatorname{Ad}_h\xi+\eta,\rho^*_h(a)),b\right)
 =\frac{d}{dt}\Big|_0\ell(\operatorname{Ad}_h\xi+\eta,\rho^*_h(a)+tb)
=\frac{d}{dt}\Big|_0\ell(\xi,a+t\rho^*_{h^{-1}}(b))
\\
\hphantom{\left(\frac{\delta\ell}{\delta a}(\operatorname{Ad}_h\xi+\eta,\rho^*_h(a)),b\right)}{}
 =\left(\rho_{h^{-1}}\frac{\delta\ell}{\delta a}(\xi,a),b\right)
\end{gather*}
for all $b\in V^*$.
\end{proof}

The path group $C^{\infty}(I,H)$ acts on $C^{\infty}(I,\mathfrak g\times V^*)$~by
\begin{gather}
\label{linact}
h\cdot (\xi,a)=(\operatorname{Ad}_h\xi+\delta^rh,\rho^*_h(a)).
\end{gather}
This action has the property $h\cdot(\delta^rg,\rho^*_ga)=(\delta^r(hg),\rho^*_{hg}{a})$ for any curve $g\in
C^{\infty}(I,G)$.

\begin{Proposition}
\label{thmlinear}
Given a~reduced Lagrangian $\ell:\mathfrak g\times V^*\to\mathbb R$ that has the invariance pro\-per\-ty~\eqref{inv}, the EP
equation with advected parameters~\eqref{rightep} is $C^{\infty}(I,H)$-invariant for the action~\eqref{linact}.
\end{Proposition}

\begin{proof}
We need the $G$-equivariance of the diamond operation:
\begin{gather*}
\operatorname{Ad}^*_g(v\diamond\rho^*_g(a))=\rho_g(v)\diamond a
\end{gather*}
that follows from $\rho^*_g(\xi a)=(\operatorname{Ad}_g\xi)(\rho^*_ga)$.
Using also the following identities for $\alpha\in\mathfrak g^*$:
\begin{gather*}
\operatorname{ad}^*_{\operatorname{Ad}_h\xi}\operatorname{Ad}^*_{h^{-1}}\alpha=\operatorname{Ad}^*_{h^{-1}}\operatorname{ad}^*_\xi\alpha,
\qquad
\frac{d}{dt}(\operatorname{Ad}^*_{h^{-1}}\alpha) =-\operatorname{ad}^*_{\delta^rh}\operatorname{Ad}^*_{h^{-1}}\alpha,
\end{gather*}
we compute
\begin{gather*}
\left(\frac{d}{dt}\frac{\delta\ell}{\delta\xi}+\operatorname{ad}^*_\xi\frac{\delta
\ell}{\delta\xi}-\frac{\delta\ell}{\delta a}\diamond a\right)(h\cdot(\xi,a))
\\
\qquad
=\frac{d}{dt}\left(\operatorname{Ad}^*_{h^{-1}}\frac{\delta \ell}{\delta\xi}(\xi,a)\right)
+\operatorname{ad}^*_{\operatorname{Ad}_h\xi+\delta^rh}\operatorname{Ad}^*_{h^{-1}}\frac{\delta\ell}{\delta\xi}(\xi,a)
-\rho_{h^{-1}}\frac{\delta\ell}{\delta a}(\xi,a)\diamond\rho^*_ha
\\
\qquad
=\operatorname{Ad}^*_{h^{-1}}\left(\frac{d}{dt}\frac{\delta\ell}{\delta\xi}+\operatorname{ad}^*_\xi\frac{\delta
\ell}{\delta\xi}-\frac{\delta\ell}{\delta a}\diamond a\right)(\xi,a).
\end{gather*}
This ensures the $C^{\infty}(I,H)$-invariance of the EP equation with advected parameters.
\end{proof}

\begin{Example}
Let~$M$ be a~Riemannian manifold.
As in Example~\ref{ex:HS} we focus on the group of volume preserving dif\/feomorphisms
$\operatorname{Dif\/f}_{\operatorname{vol}}(M)$ and the homogeneous space
$\operatorname{Dif\/f}(M)/\operatorname{Dif\/f}_{\operatorname{vol}}(M)$ of volume forms with constant total volume.

We consider the parameter space $C^{\infty}(M)^*$, identif\/ied with $C^{\infty}(M)$ via the volume form~$\mu$, hence the
left $\operatorname{Dif\/f}(M)$-action and its inf\/initesimal $\mathfrak X(M)$-action are
\begin{gather*}
\varphi\cdot\rho=\big(\rho\circ\varphi^{-1}\big)\operatorname{Jac}(\varphi^{-1}),
\qquad
u\rho =-L_u\rho-\rho\operatorname{div} u=-\operatorname{div}(\rho u).
\end{gather*}
The diamond operation~\eqref{dia} becomes
\begin{gather*}
\diamond: \ C^{\infty}(M)\times C^{\infty}(M)^*\to\mathfrak X(M)^*,
\qquad
f\diamond\rho=\rho df\in\mathfrak X(M)^*,
\end{gather*}
where the dual of the space of vector f\/ields is identif\/ied via the volume form~$\mu$ with the space of dif\/ferential 1-forms.

The reduced Lagrangian $\ell:\mathfrak X(M)\times C^{\infty}(M)^*\to\mathbb R$ given~by
\begin{gather*}
\ell(u,\rho)=\frac12\int_M\rho(\operatorname{div} u)^2\mu
\end{gather*}
comes from a~$\operatorname{Dif\/f}(M)$-invariant parameter dependent Lagrangian on
$\operatorname{Dif\/f}(M)/\operatorname{Dif\/f}_{\operatorname{vol}}(M)$.
Indeed, it satisf\/ies the invariance property~\eqref{inv}: for all $\psi\in\operatorname{Dif\/f}_{\operatorname{vol}}(M)$
and $w\in\mathfrak X_{\operatorname{vol}}(M)$ we compute
\begin{gather*}
\ell(\operatorname{Ad}_\psi u+w,\psi\cdot\rho)
=\frac12\int_M\big(\big(\psi^{-1}\big)^*\rho\big)\big(\operatorname{div}(\psi^{-1})^*u\big)^2\mu
=\frac12\int_M\rho(\operatorname{div} u)^2\psi^*\mu=\ell(u,\rho),
\end{gather*}
using at step 2 the identity $\operatorname{div}(\psi^*u)=\psi^*\operatorname{div} u$ that holds for any volume
preserving dif\/feo\-mor\-phism~$\psi$.

The EP equation with advected parameters~\eqref{rightep} becomes
\begin{gather*}
%\label{GHS}
\partial_t d(\rho\operatorname{div} u)+d L_u (\rho\operatorname{div} u)+ d\big(\rho(\operatorname{div} u)^2\big)=0,
\qquad
\partial_t\rho+\operatorname{div}(\rho u)=0.
\end{gather*}
By Proposition~\ref{thmlinear} this equation   is $C^{\infty}(I,\operatorname{Dif\/f}_{\operatorname{vol}}(M))$-invariant
for the joint action~\eqref{linact}, namely
\begin{gather*}
\psi\cdot(u,\rho)=\big(\big(\psi^{-1}\big)^* u+\delta^r\psi,\big(\psi^{-1}\big)^*(\rho)\operatorname{Jac}\big(\psi^{-1}\big)\big)
\end{gather*}
for curves~$\psi$ in $\operatorname{Dif\/f}_{\operatorname{vol}}(M)$,~$u$ in $\mathfrak X(M)$, and~$\rho$ in
$C^{\infty}(M)$.
\end{Example}

\begin{Example}
One can consider as well the group $\operatorname{Dif\/f}_{\operatorname{iso}}(M)$ of isometries of~$M$ as a~subgroup of
$\operatorname{Dif\/f}(M)$, like in Example~\ref{RM}.
The reduced Lagrangian $\ell:\mathfrak X(M)\times C^{\infty}(M)^*\to\mathbb R$ would be given~by
$\ell(u,\rho)=\frac12\int_M\rho|L_ug|^2\mu$, coming from a~$\operatorname{Dif\/f}(M)$-invariant parameter dependent
Lagrangian on the homogeneous space $\operatorname{Dif\/f}(M)/\operatorname{Dif\/f}_{\operatorname{iso}}(M)$.
\end{Example}

\section{EP equations for symmetry breaking}
\label{s5}

One can replace the linear action of~$G$ on a~parameter vector space $V^*$ with an arbitrary action of~$G$ on
a~parameter manifold~$M$.
This generalization of the EP equations with advected parameters, called EP equations for symmetry breaking, are
presented in~\cite{GBT}.
In this section we adapt these results to the case of homogeneous spaces.

\subsection{The case of Lie groups}

Let a~Lie group~$G$ act on the smooth manifold~$M$ from the left, and let $\xi_M\in\mathfrak X(M)$ denote the
inf\/initesimal generator of $\xi\in\mathfrak g$.
Given a~curve~$g$ in~$G$ starting at the identity, the curve $m(t)=g(t)\cdot m_0$ is the unique solution of the
dif\/ferential equation with time-dependent coef\/f\/icients
\begin{gather*}
\dot{m}=\xi_M(m),
\qquad
m(0)=m_0,
\end{gather*}
where $\xi(t)=\delta^rg(t)$.

The cotangent momentum map $J:T^*M\to \mathfrak g^*$, def\/ined by $(J(\alpha_m),\xi)=(\alpha_m,\xi_M(m))$ for all
$\alpha_m\in T^*_mM$, is $G$-equivariant for the cotangent and coadjoint actions:
$J(g\cdot\alpha_m)=\operatorname{Ad}^*_gJ(\alpha_m)$.

Given a~right $G$-invariant Lagrangian $L:TG\times M\rightarrow\mathbb{R}$, i.e.\
\begin{gather*}
L\big(v_gh,h^{-1}\cdot m\big)=L(v_g,m),
\qquad
h\in G,
\end{gather*}
its reduced Lagrangian {$\ell:\mathfrak{g}\times M\rightarrow\mathbb{R}$} satisf\/ies $L(v_g,m)=\ell(v_gg^{-1},g\cdot m)$.
The functional derivative $\frac{\delta\ell}{\delta\xi}$ takes values in $\mathfrak g^*$, while
$\frac{\delta\ell}{\delta m}$ is a~$\mathfrak g$-dependent section of $T^*M$.

\begin{Theorem}[\cite{GBT}]
The EL equations for the Lagrangian $L_{m_0}:TG\to\mathbb R$ are the EP equations for symmetry breaking
\begin{gather}
\label{epbreak}
{\frac{d}{dt}\frac{\delta\ell}{\delta\xi}+\operatorname{ad}^*_\xi\frac{\delta
\ell}{\delta\xi}=J\left(\frac{\delta\ell}{\delta m}\right)},
\qquad
\dot{m}=\xi_M(m)
\end{gather}
for the reduced Lagrangian $\ell:\mathfrak g\times M\to\mathbb R$.
\end{Theorem}

\begin{Example}[\cite{GBT}]
\label{nema}
For an EP description of nematic particles one considers the canonical action of $G=\operatorname{SO}(3)$ on $M=\mathbb
R P^2$.
The $\operatorname{SO}(3)$-invariant Lagrangian is
\begin{gather*}
L: \ T\operatorname{SO}(3)\times\mathbb R P^2\to\mathbb R,
\qquad
L(g,\dot g,m)=\frac12 j|\dot g|^2- \frac{\lambda}{2}\big\langle m, g^{-1}k\big\rangle^2,
\end{gather*}
where~$j$ and~$\lambda$ are constants, and~$k$ the external force f\/ield, with reduced Lagrangian
$\ell:\mathfrak{so}(3)\times\mathbb R P^2\to\mathbb R$ given by $\ell(\xi,m)=\frac12 j|\xi|^2-\frac{\lambda}{2}\langle
m, k\rangle^2$.
\end{Example}

\subsection{The case of homogeneous spaces}

Let $L:TG\times M\to\mathbb R$ be the pull-back of the right $G$-invariant Lagrangian $\bar L:T(G/H)\times M\to\mathbb
R$, hence~$L$ is left $TH$-invariant (in the f\/irst argument) and right $G$-invariant (in both arguments simultaneously).
The associated reduced Lagrangian $\ell:\mathfrak g\times M\to\mathbb R$ is both~$H$- and $\mathfrak h$-invariant:
\begin{gather}
\label{invinvinv}
\ell(\operatorname{Ad}_h\xi+\eta,h\cdot m)=\ell(\xi,m),
\qquad
h\in H,
\qquad
\eta\in\mathfrak h.
\end{gather}

\begin{Proposition}
\label{thmb}
Given a~reduced Lagrangian $\ell:\mathfrak g\times M\to\mathbb R$ that has the invariance pro\-per\-ty~\eqref{invinvinv},
the EP equation for symmetry breaking~\eqref{epbreak} is invariant under the $C^{\infty}(I,H)$-action on
$C^{\infty}(I,\mathfrak g\times T^*M)$:
\begin{gather*}
%\label{bact}
h\cdot (\xi,m)=(\operatorname{Ad}_h\xi+\delta^rh,{h}\cdot m).
\end{gather*}
\end{Proposition}

\begin{proof}
The equivariance property of the functional derivative $\frac{\delta\ell}{\delta\xi}$ from Lemma~\ref{barbar} holds, but
also the following equivariance property of $\frac{\delta\ell}{\delta m}$:
\begin{gather*}
\frac{\delta\ell}{\delta m}(h\cdot(\xi, m))={h^{-1}}\cdot\frac{\delta \ell}{\delta m}(\xi,m).
\end{gather*}
Indeed, for any curve~$c$ in~$M$ with $c(0)=h\cdot m$ and $c'(0)=w$, we get:
\begin{gather*}
\left(\frac{\delta\ell}{\delta m}(h\cdot(\xi,m)),w\right)
=\frac{d}{dt}\Big|_0\ell(\operatorname{Ad}_h\xi+\delta^rh,c(t)) =\frac{d}{dt}\Big|_0\ell(\xi,h^{-1}\cdot c(t))
\\
\hphantom{\left(\frac{\delta\ell}{\delta m}(h\cdot(\xi,m)),w\right)}{}
=\left(\frac{\delta\ell}{\delta m}(\xi,m),h^{-1}\cdot w\right) =\left(h^{-1}\cdot\frac{\delta\ell}{\delta
m}(\xi,m),w\right).
\end{gather*}
Using also the equivariance of the cotangent momentum map, we compute
\begin{gather*}
\left(\frac{d}{dt}\frac{\delta\ell}{\delta\xi}+\operatorname{ad}^*_\xi\frac{\delta
\ell}{\delta\xi}-J\left(\frac{\delta\ell}{\delta m}\right)\right)(h\cdot (\xi,m))
 =\frac{d}{dt}\left(\operatorname{Ad}^*_{h^{-1}}\frac{\delta \ell}{\delta\xi}\right)
+\operatorname{ad}^*_{\operatorname{Ad}_h\xi+\delta^rh}\operatorname{Ad}^*_{h^{-1}}\frac{\delta\ell}{\delta\xi}
\\
\qquad
{}-J\left(h^{-1}\cdot\frac{\delta\ell}{\delta
m}\right) =\operatorname{Ad}^*_{h^{-1}}\left(\frac{d}{dt}\frac{\delta\ell}{\delta\xi}+\operatorname{ad}^*_\xi\frac{\delta
\ell}{\delta\xi}-J\left(\frac{\delta\ell}{\delta m}\right)\right).
\end{gather*}
This shows the required invariance of the equation~\eqref{epbreak}.
\end{proof}

\begin{Example}
This is a~variation of Example~\ref{nema} for the subgroup $H=S^1$ of $G=\operatorname{SO}(3)$ consisting of all
rotations with axis~$k$.
Instead of the reduced Lagrangian $\ell(\xi,m)=\frac12 j|\xi|^2-\frac{\lambda}2(m\cdot k)^2$ from Example~\ref{nema} one
can take $\ell(\xi,m)=\frac12 j|p_{k^\perp}(\xi)|^2-\frac{\lambda}2 \langle m, k\rangle^2$, where $p_{k^\perp}$ denotes
the orthogonal Euclidean projection onto the vectorial plane $k^\perp\subset\mathbb R^3$.
It has the required invariance properties because $p_{k^\perp}(\eta)=0$ for all $\eta\in\mathfrak h$ (because~$\eta$ is
proportional to~$k$) and $p_{k^\perp}(\operatorname{Ad}_h\xi)=\operatorname{Ad}_{h}(p_{k^\perp}(\xi))$ for all $h\in H$.
Indeed, for every rotation~$h$ with axis~$k$ and every $\eta\in\mathfrak h$,
\begin{gather*}
\ell(\operatorname{Ad}_h\xi+\eta,h\cdot m)=\frac12 j|p_{k^\perp}(\operatorname{Ad}_h\xi)|^2-\frac{\lambda}{2}\langle
h\cdot m,k\rangle^2=\frac12 j|p_{k^\perp}(\xi)|^2-\frac{\lambda}{2}\langle m,k\rangle^2=\ell(\xi,m).
\end{gather*}
\end{Example}

\subsection{Af\/f\/ine EP equations}
%\label{s4}

Now we consider the special case of an af\/f\/ine left $G$-action on a~parameter space $V^*$:
\begin{gather}
\label{aff}
\theta_g(a)=\rho^*_{g}(a)+c(g),
\end{gather}
where $c:G\to V^*$ is a~group 1-cocycle for the action $\rho^*$, i.e.\
\begin{gather*}
%\label{cocycle}
c(gh)= c(g)+\rho^*_{g} c(h).
\end{gather*}
Let $dc:\mathfrak g\to V^*$ be the associated Lie algebra 1-cocycle.
If $\xi(t)=\delta^rg(t)$, then $a(t)=\theta_{g(t)}(a_0)$ is the unique solution of the dif\/ferential equation with
time-dependent coef\/f\/icients $\dot{a}=\xi a+dc(\xi)$, $ a(0)=a_0$.

\begin{Remark}
Let $dc^\top:V\to\mathfrak g^*$ be def\/ined by $\langle dc^\top(v),\xi\rangle=\langle dc(\xi),v\rangle$.
Then the cotangent momentum map for the af\/f\/ine action~\eqref{aff} of~$G$ on $V^*$ can be written as
\begin{gather}
\label{cota}
J: \ T^*V^*=V^*\times V\to\mathfrak g^*,
\qquad
J(a,v)=v\diamond a+dc^\top(v),
\end{gather}
because for all $\xi\in\mathfrak g$,
\begin{gather*}
(J(a,v),\xi)=(v,\xi_{V^*}(a))=(v,\xi a+dc(\xi))=\big(v\diamond a+dc^\top(v),\xi\big).
\end{gather*}
\end{Remark}

The following result for a~right $G$-invariant Lagrangian $L:TG\times V^*\rightarrow\mathbb{R}$ with reduced Lagrangian
$\ell:\mathfrak{g}\times V^*\rightarrow\mathbb{R}$ is a~special case of Theorem~\ref{epbreak} and a~generalization of
Theorem~\ref{lini}.

\begin{Theorem}[\cite{GBR}]
The EL equations for $L_{a_0}:TG\to\mathbb R$ can be expressed as affine EP equations for the reduced
Lagrangian~$\ell$:
\begin{gather}
\label{epaffine}
{\frac{d}{dt}\frac{\delta\ell}{\delta\xi}+\operatorname{ad}^*_\xi\frac{\delta \ell}{\delta\xi}=\frac{\delta\ell}{\delta
a}\diamond a+(dc)^\top\left(\frac{\delta \ell}{\delta a}\right)},
\qquad
\dot{a}=\xi a+dc(\xi).
\end{gather}
\end{Theorem}

{\bf Spin systems.} In~\cite{GBR} is shown that the af\/f\/ine EP equations for the action of the gauge group
$C^{\infty}(M,G)$ on the space $V^*=\Omega^1(M,\mathfrak g)$ of principal connections on the trivial bundle $M\times G$
\begin{gather}
\label{gaug}
\theta_g(\gamma)=\operatorname{Ad}_{g}\gamma-dgg^{-1}
\end{gather}
can be used in the description of spin systems.
The 1-cocycle is in this case the right logarithmic derivative
\begin{gather*}
%\label{cocy}
c: \ C^{\infty}(M,G)\to\Omega^1(M,\mathfrak g),
\qquad
c(g)=-dgg^{-1},
\end{gather*}
so $dc(\xi)=-d\xi$ for all $\xi\in C^{\infty}(M,\mathfrak g)$.
The inf\/initesimal action involves the covariant derivative $d^\gamma\xi=d\xi+[\gamma,\xi]$, namely
$\xi_{V^*}(\gamma)=-d^\gamma\xi$.

We f\/ix a~volume form on~$M$, so $C^{\infty}(M,\mathfrak g^*)$ is a~dual space to the gauge Lie algebra
$C^{\infty}(M,\mathfrak g)$, while $\mathfrak X(M,\mathfrak g^*)$ is a~dual space to the parameter space
$\Omega^1(M,\mathfrak g)$.
The cotangent momentum map~\eqref{cota} becomes $J(\gamma,\alpha)
=-\operatorname{ad}^*_\gamma\alpha-\operatorname{div}\alpha=\operatorname{div}^\gamma\alpha$, since
$dc^\top(\alpha)=\operatorname{div}\alpha$ and that the diamond map is
$\alpha\diamond\gamma=-\operatorname{ad}^*_\gamma\alpha$.
We can write now the af\/f\/ine EP equation on $C^{\infty}(M,\mathfrak g)\times\Omega^1(M,\mathfrak g)$ as
\begin{gather}
\label{spin}
\frac{\partial}{\partial
t}\frac{\delta\ell}{\delta\xi}+\operatorname{ad}^*_\xi\frac{\delta\ell}{\delta\xi}=-\operatorname{div}^\gamma\frac{\delta\ell}{\delta
\gamma},
\qquad
\dot\gamma+d^\gamma\xi=0.
\end{gather}
For $M=\mathbb R^3$ and $G=\operatorname{SO}(3)$ one gets a~macroscopic description of spin glasses~\cite{GBR}.
For~$M$ a~real interval and $G=\operatorname{SE}(3)$, the Euclidean group of rigid motions, one gets an af\/f\/ine EP
formulation of Kirchhof\/f's theory of rods (the Cosserat rod) in the case of potential forces~\cite{GBHR}.

{\bf Homogeneous spaces.} Let $L:TG\times V^*\to\mathbb R$ be now the pull-back of a $G$-invariant Lagrangian $\bar
L:T(G/H)\times V^*\to\mathbb R$.
Because~$L$ is left $TH$-invariant and right $G$-invariant, its reduced Lagrangian $\ell:\mathfrak g\times V^*\to\mathbb
R$ is both~$H$- and $\mathfrak h$-invariant:
\begin{gather}
\label{invinv}
\ell(\operatorname{Ad}_h\xi+\eta,\theta_{h}(a))=\ell(\xi,a),
\qquad
h\in H,
\qquad
\eta\in\mathfrak h.
\end{gather}

\begin{Proposition}
%\label{thmaffine}
Given a~reduced Lagrangian $\ell:\mathfrak g\times V^*\to\mathbb R$ that has the invariance pro\-per\-ty~\eqref{invinv}, the
affine EP equation~\eqref{epaffine} is invariant under the action of the path group $C^{\infty}(I,H)$ on
$C^{\infty}(I,\mathfrak g\times V^*)$~by
\begin{gather}
\label{affact}
h\cdot (\xi,a)=(\operatorname{Ad}_h\xi+\delta^rh,\theta_{h}(a)).
\end{gather}
\end{Proposition}

\begin{proof}
It is a~consequence of Proposition~\ref{thmb}, but it can be shown also directly, as in the proof of
Proposition~\ref{thmlinear}, using the expression of the failure of $dc$ to be $G$-equivariant:
$dc(\operatorname{Ad}_g\xi)-\rho^*_gdc(\xi)=c(g)\operatorname{Ad}_g\xi$.
\end{proof}

\begin{Example}[spin systems]
Let~$G$ be a~Lie group and~$\kappa$ an invariant inner product on its Lie algebra $\mathfrak g$.
The reduced Lagrangians $\ell:C^{\infty}(M,\mathfrak g)\times\Omega^1(M,\mathfrak g)\to\mathbb R$ that depend only on
the dif\/ferential of the function $\xi\in C^{\infty}(M,\mathfrak g)$:
\begin{gather*}
\ell_1(\xi,\gamma)=\frac12\int|[d\xi,\gamma]|^2\mu,
\qquad
\ell_2(\xi,\gamma)=\frac12\int|\kappa(d\xi,\gamma)|^2\mu,
\qquad
\ell_3(\xi,\gamma)=\int\big(|d\xi|^2-|\gamma|^2\big)\mu
\end{gather*}
all come from a~Lagrangian on the homogeneous space $C^{\infty}(M,G)/G$ because all of them satisfy the invariance
property~\eqref{invinv}.
We check it for the middle Lagrangian for all $\xi\in C^{\infty}(M,\mathfrak g)$, $h\in G$ and $\eta\in\mathfrak g$ (so
$dhh^{-1}=0$ and $d\eta=0$):
\begin{gather*}
\ell_2(\operatorname{Ad}_h\xi+\eta,\theta_h(\gamma))
\stackrel{\eqref{gaug}}{=}
\frac12\!\int\!\big|\kappa\big(\operatorname{Ad}_hd\xi+d\eta,\operatorname{Ad}_h\gamma-dhh^{-1}\big)\big|^2\mu
=|\kappa(d\xi,\gamma)|^2\mu=\ell_2(\xi,\gamma).
\end{gather*}
It follows that the corresponding EP equations~\eqref{spin} for spin systems are invariant under the
action~\eqref{affact}, hence~\eqref{spin} can be seen as an equation on $C^{\infty}(M,G)/G$.

This setting of af\/f\/ine EP reduction is used in~\cite{V} for the dynamical description of space-time strands on
homogeneous spaces.
Covariant EP equations on homogeneous spaces provide another frame to describe the dynamics of space-time strands on
homogeneous spaces.
\end{Example}

\subsection*{Acknowledgements}

The author is grateful to the referee for very helpful suggestions.
This work was supported by a~grant of the Romanian National Authority for Scientif\/ic Research, CNCS UEFISCDI, project
number PN-II-ID-PCE-2011-3-0921.

\vspace{-1mm}

\pdfbookmark[1]{References}{ref}
\LastPageEnding

\end{document}